# CARBON ECONOMICS OF DIFFERENT AGRICULTURAL PRACTICES FOR FARMING SOIL

Suganthi Pazhanivel Koushika[1], Anbalagan Krishnaveni[2]*, Sellaperumal Pazhanivelan[3], Alagirisamy Bharani[1], Venugopal Arunkumar[2] Perumal Devaki [1] and Narayanan Muthukrishnan [2]

[1] Department of Environmental Science, Tamil Nadu, Agricultural University, Coimbatore

[2] Department of Crop Management, TNAU, Vazhavachanur, Tiruvannamalai

[3] Centre for Water and Geospatial Studies, TNAU, Coimbatore

* Corresponding author - krishnavenia@tnau.ac.in

**Abstract**

The loss of soil organic carbon (SOC) poses a severe danger to agricultural sustainability around the World. This review examines various farming practices and their impact on soil organic carbon storage. After a careful review of the literature, most of the research indicated that different farming practices, such as organic farming, cover crops, conservation tillage, and agroforestry, play vital roles in increasing the SOC content of the soil sustainably. Root exudation from cover crops increases microbial activity and helps break down complex organic compounds into organic carbon. Conservation tillage enhances the soil structure and maintains carbon storage without disturbing the soil. Agroforestry systems boost organic carbon input and fasten nutrient cycling because the trees and crops have symbiotic relationships. Intercropping and crop rotations have a role in changing the composition of plant residues and promoting carbon storage. There were many understanding on the complex interactions between soil organic carbon dynamics and agricultural practices. Based on the study, the paper reveals, the role of different agricultural practices like Carbon storage through cover crops, crop rotation, mulching Conservation tillage, conventional tillage, zero tillage and organic amendments in organic carbon storage in the soil for maximum crop yield to improve the economical condition of the cultivators.

## 1. Introduction

Different land use patterns have resulted in a 31 per cent or 270 ± 30 Penta gram (Pg) rise in atmospheric carbon content since the Industrial Revolution. The carbon content of up to 78 ± 12 Pg has been added to the atmosphere through soil organic matter (SOM) depletion and with a cumulative loss of 30 – 40 Mega gram (Mg) carbon or two-thirds of their initial SOC ha$^{-1}$ of agricultural soil. One of the most incredible solutions to remove atmospheric carbon and store it in soil requires implementing different farming practices. Agricultural soils are believed to be a significant carbon sink and can sequester more carbon. Based on the benefits of other farming practices for promoting agricultural sustainability through carbon storage and reducing climate change. Agroecosystems are practiced for their carbon storage (CS) potential of 12–228 Mg ha$^{-1}$. Of agricultural soil. A 1.1–2.2 Penta gramPg C can be sequestered in $585 – 1215 \times 10^6$ ha of Earth's usable area through crop production in the next 50 years(Albrecht & Kandji, 2003). Agricultural plants and soil have enormous carbon storage potential through different farming practices. Better carbon storage may also come from modifications of farm practices and management techniques. The agricultural yield is expected to decline when the management practices are not sustainable. Compelling the carbon emissions from the soil due to applying nitrogen fertilizer to the environment requires increased SOM. Maintaining agricultural soil health involves using organic manures(Schlesinger & Andrews, 2000)

Increasing the area of agricultural operations and intensive crop production techniques such as irrigation boost crop yields but deteriorate soil carbon content(Baker et al., 2007). The deterioration of the land is inevitable by intensive agricultural management practices. Sustainable intensification can enhance farm yield and reduce the adverse environmental effects on agrarian soil (Kucharik et al., 2001). Soil organic matter (SOM) loss is one of the indicators of land degradation. SOM influences several characteristics of soils, such as water-holding capacity and nutrient stability, to provide structure for adequate drainage and aeration. It also prevents topsoil loss due to erosion by rebuilding SOM(Reeves, 1997). One of the essential aspects of sustainable intensification is storing SOM, which ensures a sustainable crop yield and less dependency on chemical fertilizers. Nevertheless, the focus on soil management has been qualitative. Still, it is unclear how much building soil organic matter will contribute to boosting crop productivity and reducing agriculture's adverse environmental effects(Adhikari & Hartemink, 2016).

**2. Role of soil organic carbon**

Based on the category of soil organic matter, the soil organic carbon is measurable in different agricultural soils. Farm soils depend on organic matter for their physical, chemical, and biological functions, even though they contribute 2–10 % of the bulk of the soil. In addition to aiding in carbon storage, organic matter helps with soil structure, nutrient retention and turnover, moisture availability and retention, and pollutant degradation. To lower atmospheric carbon dioxide, sequestering carbon is essential for the climate change mitigation strategy. The increased soil organic carbon in agricultural and pastoral lands will considerably reduce atmospheric carbon dioxide(Griffin et al., 2013)**.** The most significant environmental problem of the twenty-first century is the need to stabilize the atmosphere's greenhouse gas concentration. Humanity can either find ways to remove greenhouse gases once they have been released from the atmosphere or lower the emissions of fossil fuels to manage their concentrations. The magnitude of the relevant fluxes makes the topic difficult to tackle. Currently, about 10 Giga tons of carbon is released into the atmosphere each year by industry, transportation, and residential usage, and there is no imminent prospect of a significant decline in these emission rates(Jackson, 2017)**.**

One alternative strategy to slow the pace of the increase of greenhouse gases and the corresponding changes in our climate is the storage of carbon dioxide from the atmosphere, which acts as organic carbon in the biosphere(Amundson, 2018). Science experts have assessed the possibility of sequestering carbon in soil organic matter for almost 20 years(Lal, 2001). The idea is logical in and of itself: over 10,000 years of agricultural cultivation have resulted in a 116 Giga tons reduction in worldwide soil carbon, which is more than ten times the current rate of industrial emissions. It is suggested that by altering agricultural practices, a large portion of this carbon may be returned to cultivated soils, acting as a significant instrument to slow down global warming and give humanity more time to decarbonizes(Sanderman et al., 2017) and at the same time carbon serves the following purposes in the life of plants: Carbon is fixed in biological form by plants and other organisms called producers through photosynthesis, which converts solar energy into chemical form. The most fundamental of all sugars, glucose, is created during photosynthesis with the help of light energy, carbon dioxide, and water. Afterward, the carbohydrates are converted into the chemical energy that drives the cells of all plants and animals. For longer-term energy storage, some plant carbon is converted into large, complex molecules like starch, while other carbon remains as simple glucose for immediate energy use.

## 3. Carbon storage through different agricultural practices

### *3.1 Conservation tillage, conventional tillage and zero tillage*

Many regions now view conservation agriculture (CA) as a viable substitute for conventional agriculture. It is said to be a productive way to enhance soil quality and crop performance, and it has a beneficial impact on mitigating climate change(Kassam & Friedrich, 2012; Zhang, 2022). A different approach to raising agricultural output sustainably has been mentioned: the conservation agriculture (CA) system. This approach is generally accepted to increase infiltration rates, lessen erosion issues, and boost the quality of soil and organic carbon levels in agricultural environments(Kahlon et al., 2013; Lal, 2001). The practice of conservation agriculture, which involves the use of woody crops and residue-based zero tillage, has several benefits, such as enhancing soil aggregation, minimizing soil compaction, preventing soil erosion, lowering weed infestations, improving water infiltration into deeper soil layers, lowering production costs, and maintaining some fallow land through direct seeding(Giller et al., 2009). This has recently attracted much attention due to worries about growing atmospheric carbon dioxide levels. An estimated 1500 Giga tons of SOC, or about twice as much carbon as in the atmosphere, are found in the world's soils(Schlesinger, 2000). Estimates of carbon lost from US croplands (an average of 36 tons per hectare) range from 5 Giga tons. These estimates blame soil plowing since many consider tilled soils depleted carbon reservoirs that can be replenished. With proper management, there is hope to restore much of this over 50 years(Larney & Kladivko, 1989).

It has been suggested that the United States could store between 24 and 40 million tons of carbon annually if conservation tillage were widely adopted. One of the most critical worldwide methods for stabilizing atmospheric $CO_2$ concentrations is the conversion of all croplands to conservation tillage, which is expected to absorb 25 Giga tons of carbon over the next 50 years(Pacala & Socolow, 2004). Because soil tillage influences both aggravating and degrading processes, it impacts SOC. Crop residue and other biomass are humified, increasing resistant or non-labile fraction of soil organic carbon content; SOC is sequestered in creating organo-mineral complexes and increases stable aggregation; and SOC is deeply deposited in subsoil layers. These are some of the aggravating processes of soil that improve SOC(Dexter, 1991).

In contrast, soil-degrading processes such as leaching, mineralization, and erosion hurt SOC. The term "conservation tillage" refers to a broad range of approaches that use crop residual mulch as a raindrop screen while reducing soil and water losses compared to standard or plow-based tillage. These strategies boost soil organic matter (SOC) by hastening soil aggravating and counteracting degrading processes(Carter, 2017). Maintaining crop residues on the surface and zero tillage (ZT) has been demonstrated to successfully prevent soil disintegration, improve soil fertility, and boost the capacity for SOC storage to adapt to and buffer the effects of global warming(Nath et al., 2017). Still, the amount and caliber of crop leftover mulch reapplied to the field determines its success(Fonseca et al., 2022).

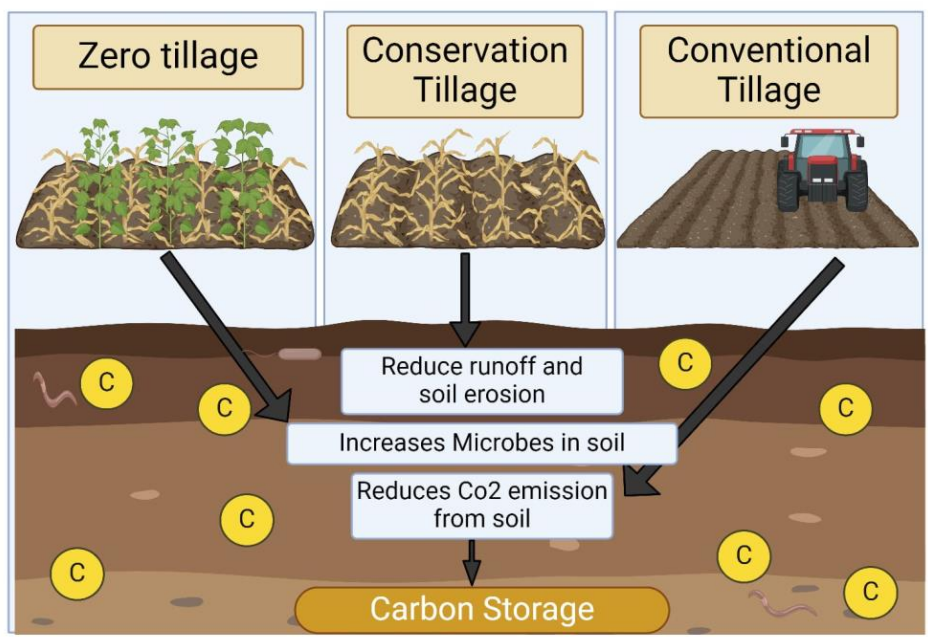

**Fig.1 Carbon storage under zero tillage, conservation tillage, and conventional tillage**

Apart from soil depth, other factors that influence soil carbon storage rates are soil properties (such as clay content and type, fertility status, and soil water retention), landscape orientation (less in shoulder and summit slopes than foot slopes), profile and terrain characteristics (more in young soils with deep adequate rooting depth and less in south-facing and convex slopes than in north-facing and concave slopes), and climatic conditions (less in warm than cold climates and less in dry than humid regions). By switching from conventional to conservation tillage, a significant emphasis was placed on the amount and range of SOC storage(Kimble et al., 1998).Long-term use of residue-based zero tillage is a successful strategy for storing atmospheric $CO_2$ in the soil, which preserves high agricultural yields while protecting natural resources(Nath et al., 2017; Sa & Lal, 2009) Additionally, it has been shown that maintaining crop residues on the surface along with zero tillage (ZT) can effectively reduce soil disintegration, improve soil fertility, and increase soil organic carbon storage potentials for adaptation and mitigation of the effects of global warming(Wang et al., 2020). Even though it eventually causes a rise in soil carbon, the evidence shown thus far is weak. Additional studies, including deeper soil samples and long-term gas exchange studies, might clarify this issue. Until then, it is too soon to estimate how much carbon farming systems could sequester based on anticipated changes in tillage techniques or to encourage such changes through carbon sequestering policies or market tools. While trying to encourage the adoption of alternative land use and management strategies that may be shown to reduce the increase in greenhouse gas concentrations in the atmosphere, the scientific community runs the danger of losing credibility(Sarkar & Singh, 2007). As mentioned in Fig.1 Carbon storage under zero tillage, conservation tillage, and conventional tillage" is the title of a study that compares the effects of various farming methods on soil carbon sequestration. Conventional tillage involves more soil manipulation, zero tillage requires little to no soil disturbance, and conservation tillage uses techniques to prevent soil erosion and preserve soil health.

Tillage techniques such as conservation, conventional, and zero tillage impact soil carbon storage. Conservation tillage entails reducing soil disturbance and leaving crop leftovers on the field surface to promote carbon retention in the soil. Conversely, conventional tillage causes more significant soil disruption, which can hasten the breakdown of organic materials and release carbon into the atmosphere. Zero tillage, often known as no-till farming, is a method that leaves the soil undisturbed, allowing carbon to accumulate in the soil over time. According to research, conservation and zero tillage approach boost carbon sequestration by minimizing soil erosion and organic matter breakdown, resulting in better soil structure and carbon content. As a result, implementing conservation or zero tillage methods may significantly increase carbon storage in agricultural soils, playing an essential role in climate change mitigation and sustainable land management.

### *3.2 Carbon storage through cover crops, crop rotation, mulching*

Cover crops are cultivated to increase soil fertility, stop erosion, enrich and protect the soil, and improve water and nutrient availability. Several advantages cover crops offer to soils utilized in agricultural production. These crops are carefully cultivated to minimize soil erosion and stop nutrient leaching and surface flow from deep layers. Legumes are the principal component of cover crops, produced to cover the soil's surface and enhance its physical, chemical, and biological properties. Cover crops are seeded between primary crops to boost agricultural production and productivity. The perfect cover crop should be fast to germinate and emerge, resilient to harsh weather, capable of fixing atmospheric nitrogen from the air, able to absorb nutrients from the soil by growing deep roots, and able to produce a more significant amount of biomass in a shorter amount of time, easy to work and cultivate, not compete with the primary crop, resistant to diseases and insect pests, able to suppress weeds, and inexpensive to cultivate(Bayer et al., 2000; Deb et al., 2013).

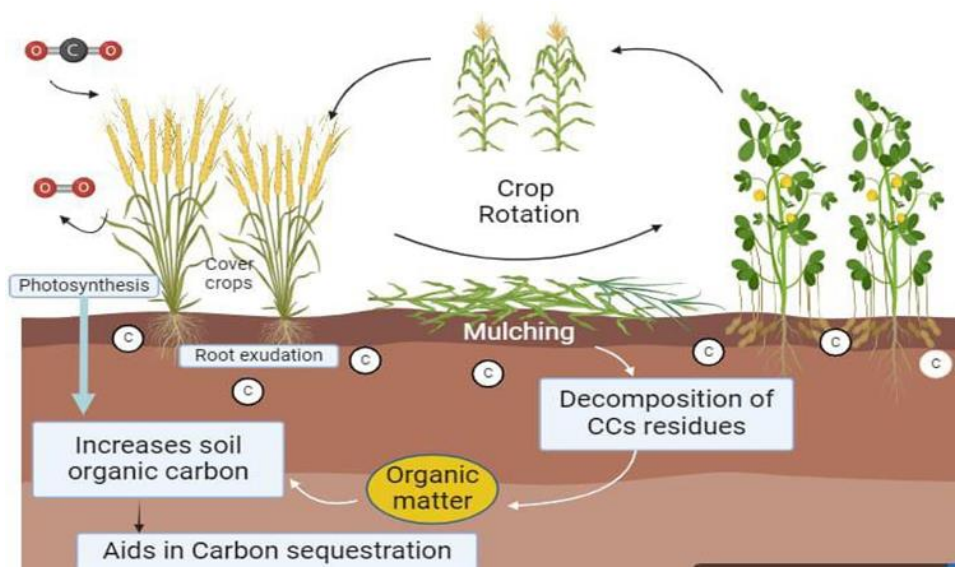

**Fig.1 Carbon storage under mulching cover crop and crop rotation**

The advantages of cover crops for the agricultural community and environment have been widely recognized for many years. Agricultural soils have a lower SOC content than soils with a natural plant cover. SOC losses from crop cultivation are 30–40% higher than those from wild vegetation (Don et al., 2011; Poeplau et al., 2011; Shirley & Teasdale, 1998; Sohrabi et al., 2012). Due to the variations in plant carbon inputs and rates of mineralization, different crop management techniques can impact the amount of soil organic carbon sequestered in conventional tillage and no-till soils. In a study at Fort Valley State University, it was discovered that in cotton (*Gossipium spp.*) and sorghum (*Sorghum bicolour L.*), the soil organic carbon content (SOC) at 0–10 cm varied with plant carbon input and was higher from the cover crops (hairy vetch, rye, mixture of hairy vetch and rye) than from no cover crops (weeds). When cover crops were used instead of no cover crop treatment, the SOC at 0–30 cm depth increased to 120–130 kg N ha$^{-1}$yr$^{-1}$. The impact of cover crops on SOC was not considered in a German field investigation. Compared to alternative management techniques, cover crops have the benefit of having higher SOC without lowering primary crop production. The carbon response function, which describes variations in SOC over time, was evaluated using a meta-analysis of 139 plots at 37 distinct locations. Cover crops under rotation for up to 54 years showed a linear correlation with yearly SOC change at a depth of 22 cm, at a rate of $0.32 \pm 0.08$ Mg ha$^{-1}$ yr$^{-1}$. The premise that the observed linear SOC accumulation will not continue to rise indefinitely informed the average SOC stock change modeling. After 155 years of employing cover crops, the updated estimated steady state data would have a SOC buildup of $16.7 \pm 1.5$ Mg ha$^{-1}$ yr$^{-1}$(Sainju et al., 2006; Uhlen & Tveitnes, 1995).

Crop rotation is defined as a deliberate series of crops cultivated on the same land in a regularly occurring succession as opposed to continuous monoculture or farming a varied sequence of crops. Many factors, including changes in SOC content, soil aggregation and structure, nutrient cycling, and insect incidence, are impacted by crop rotation, which also affects the growth and productivity of subsequent crops. Combined with no-till or low-till farming practices, crop rotations can reduce soil erosion, boost organic carbon, and sequester soil carbon. Soil carbon addition is mainly determined by crop selection and cropping practices(Ingram & Fernandez, 2001)**.** While extensive monoculture reduces crop production and thus lowers the biomass returned to the soil, crop rotation is more successful than monoculture at retaining C and N. Crop rotation returns wastes to the soil in varying amounts, which determines the amount of SOC sequestered(Biederbeck, 1984)**.** Compared to grain-legume rotations, there was increased SOC storage due to the high residue-producing continuous sorghum in the first rotation and continuous corn in the second, as well as lower tillage and surface residue management.

**Table 1:Carbon storage rate under different crops grown in the field**

| Crop | Carbon storage rate | References |
|---|---|---|
| Barley | 0.247kg C ha$^{-1}$ | |
| Cassava | 0.5032kg C ha$^{-1}$ | |

| Rice | 0.401247 kg C ha$^{-1}$ | (Jarecki & Lal, 2003) |
|------|------------------------|------------------------|
| Corn | 0.292 | |
| Millet | 0.071 | |

The adoption of no-till management and the intensification of the cropping system from spring wheat monoculture to annual cropping rotation consisting of spring wheat, winter wheat, and sunflower had a beneficial effect on lowering the loss of soil organic carbon from croplands in the northern Great Plains(Omay, 1997). In an extended investigation, the grass pasture system yielded the maximum soil organic carbon retention. In contrast, the wheat-fallow system markedly decreased soil microbial biomass and organic carbon. Tillage, the antecedent SOC pool, and the soil characteristics modify the rotation impacts on SOC storage(Halvorson et al., 1999; Halvorson et al., 2002). The potential for SOC storage for various crop rotations and tillage management, as well as the time during which SOC storage may occur, using the worldwide database of 67 long-term agricultural trials. Because no-till treatments were used to recalculate the data, the rate of SOC storage under varying crop rotation is comparatively low. They concluded that SOC augmentation brought about by implementing a more intricate crop rotation may attain a new equilibrium in around 40 - 60 years(Collins, 1992).

Mulching also enhances the soil environment and regulates soil temperature under the NT system without negatively affecting crop productivity (Anikwe et al., 2007; Kahlon et al., 2013; Sarkar & Singh, 2007). Moreover, mulching speeds up soil erosion during heavy rains by increasing soil porosity and water-holding capacity as well as lowering surface runoff (Ab & Kulig, 2008; Bhatt & Khera, 2006; Gajri et al., 1994; Jastrow, 1996). The fresh residue is incorporated into the soil to increase the creation of microbial-imitative binding agents and to give soil microbial populations a suitable carbon source, facilitating the development of macro aggregates. According to reports, implementing no tillage decreases the macro-aggregate turnover rate compared to plough tillage, enhancing aggregate stability. Conservation tillage approaches are comparatively more successful and may be applied soon compared to other atmospheric drawdown measures. The documented advantages of enhancing soil quality and sequestering carbon are extensive, but the dangers in this system are negligible(Six et al., 2000). Let us discuss the different crops and their role in sequestering the soil carbon into the soil.Fig.2, Carbon Sequestration under mulching, cover crop, and crop rotation", probably shows how different farming techniques help to increase the amount of carbon sequestered in soil. Crop rotation, cover crops, and mulching are sustainable farming practices that improve soil health and boost carbon storage.

Carbon storage through agricultural methods such as cover crops, crop rotation, and mulching is a complex sequence of interrelated activities that substantially influence soil carbon sequestration. Cover crops are often cultivated between main crop seasons andhelp store carbon by introducing organic matter into the soil via root biomass and decomposing plant material. Crop rotation improves soil carbon storage by varying plant residues, root architectures, and nutrient needs while encouraging microbial activity and organic matter absorption. Mulching, or the application of organic or inorganic materials to the soil surface, controls soil temperature and moisture and promotes carbon retention via the progressive degradation of the mulch material. These techniques jointly increase soil structure, microbial diversity, and organic carbon content, resulting in better soil health and resilience.

### 3.3 Carbon storage in soil under the addition of organic amendments

Green manure is a crop planted to be ploughed down in the soil while still green rather than being collected for human and animal use. Green manure is widely used worldwide to maintain the health of the soil. Generally speaking, green manures are made from leguminous plants like sun hemp (*Crotalaria juncea*), pea (*Pisum sativum*), cowpea (*Vigna unguiculata*), groundnut (*Arachis hypogaea*), black Gram (*Vigna mungo*), lentil (*Lens culinaris*), clover (*Trifolium spp.*), soybean (*Glycine max*), mungbean (*Vigna radiata*), dhaincha (*Sesbania spp.*)etc. Fresh leaves of forest plants can also be used as green manure; farmers gather green leaves and twigs from the forest plants and incorporate them into the soil to increase the soil's organic matter status. This practice is known as green leaf manure. Legumes are the best type of plants to use for green manuring because they provide a significant amount of nitrogen to the soil through biological nitrogen fixation(Malaviya et al., 2019; Skoien, 1993)**.**

Organic wastes originating from food and wood resources, such as kitchen and yard wastes, are household wastes. The degradable trash produced by daily home activities is used to make compost from domestic garbage. Most household trash comprises food items, mainly the non-consumable parts of fruits and vegetables. It is essentialto manage home garbage properly to prevent pollution of the environment. Using composted household waste for crop production will make the agriculture industry less dependent on synthetic fertiliser, lowering production costs and maintaining environmental safety. Therefore, compost made from home waste products might be a desirable supply of organic matter for soil.

One significant kind of organic manure created by the breakdown of different plant and animal wastes is compost. Many waste products may be used to create high-quality compost, including plant leaves, kitchen scraps, banana and pineapple peels, weeds, water hyacinth, paper mill residue, sugarcane peels, straw, sawdust, rice husk and leftovers from animal slaughter.Trash from the leather industry, cities, and other sources can also be used to make compost. However, it must be devoid of harmful trace elements and heavy metals. For safe crop production to be guaranteed, the dangerous substances and contaminants must, at the very least, be below the threshold level. For gardeners and vegetable producers, compost is called "black gold" because of its many advantages for developing plants(Shirani, 2002)**.**

Livestock excrement, both liquid and solid, makes up "farmyard manure," an organic material typically combined with a small quantity of litter, such as straw (mostly rice straw), used to clean animals. Animal urine, leftover feed, and bedding materials comprise this byproduct. Given that it is easily obtainable and contains most of the nutrients crops need, it is one of the oldest manures used by farmers to produce various crops, particularly vegetables. The urine from cattle is a valuable component of farmyard manure because it contains much nitrogen; however, most of the pee is wasted because the animal shed's clay floor soaks up the urine. To resolve the issue, the animal shed has to be cemented. Straw, sawdust, dried weeds, rice husk, etc., might be utilised to lessen the urine from the cow shed in the case of an earthen floor(Gupta et al., 2021; Kumar et al., 2020; Shirani et al., 2002).

Several kinds of earthworms make vermicompost, a type of compost. Earthworms usually devour a variety of decomposable organic wastes; only a tiny percentage of this food - between five and ten per cent - is absorbed by the body; the majority is expelled as pellets that are then composted, or known as vermicompost. It comprises earthworm cocoons, vitamins, enzymes, beneficial microbes, auxins, gibberellins, and other hormones that promote plant development and excrement. Plants may easily absorb nutrients from vermicompost because they absorb them quickly. Adding vermicompost to the soil enhances its biological, physical, and chemical qualities(Manivannan et al., 2009).

Poultry manure is one of the significant traditional manures utilised for a long time in the agricultural industry to improve crop productivity in many places worldwide - the organic waste produced by the poultry birds excrement, urine, and bedding. Every year, more poultry birds are raised worldwide to fulfil the rising needs of an expanding population, which results in massive amounts of poultry litter. Taking good care of the chicken litter is crucial. If not, it may seriously contaminate the ecosystem. While there are many uses for chicken manure, including the production of biogas and power, fish feed, mushrooms, and so on, the most significant application is organic manure on agricultural fields(Reeves, 1997).

Organic additions increase soil organic carbon content and provide stable organic matter, which enhances soil structure, water retention, and nutrient availability. Soil bacteria decompose these organic components, producinga long-lasting and stable organic carbon humus. Increased soil organic carbon increases carbon sequestration and improves soil health and fertility. Furthermore, introducing organic amendments can boost soil microbial activity, creating a favourable environment for carbon stability. In addition to the discussed organic amendments such as green manure, green leaf manure, cow dung, and domestic waste, this supplement will improve the soil's organic matter and increase the soil's organic carbon, whichcould be helpful for crop growth(Bianchi et al., 2008).

### 3.4 Carbon storage in soil under agroforestry

An approach to land use known as agroforestry includes crops and animals with woody perennials such as trees, shrubs, palms, and bamboo. This ecologically oriented, dynamic approach to managing natural resources offers land users tremendous social, economic, and environmental advantages at all levels. Agri silvicultural systems, such as alley cropping or home gardens, mix trees and crops; silvopastoral systems, on the other hand, combine forestry with domestic animal grazing in pastures, rangelands, or on-farm(Schroeder, 1993). Over 11 per cent of India's land area, or 31.3 million hectares, was covered by natural forests in 2010. Its loss of 117 kha$^{-1}$ of natural forest in 2022 amounted to 62.9 Mt of$CO_2$ emissions**.** One quick way to reduce greenhouse gas emissions is to manage trees in agroecosystems like agroforestry, trees outside of forests, and other anthropogenically managed forests. By promoting the growth of trees and bushes, agroforestry techniques can store carbon and absorb carbon dioxide from the atmosphere. This is a viable method of sequestering carbon(Singh, 2000)**.** Agroforestry techniques have a long history in India.

In India, agroforestry systems consist of various local forest management, ethno-forestry methods, and trees planted on farms and community forestry(Dixon, 1995; Pandey, 2007).

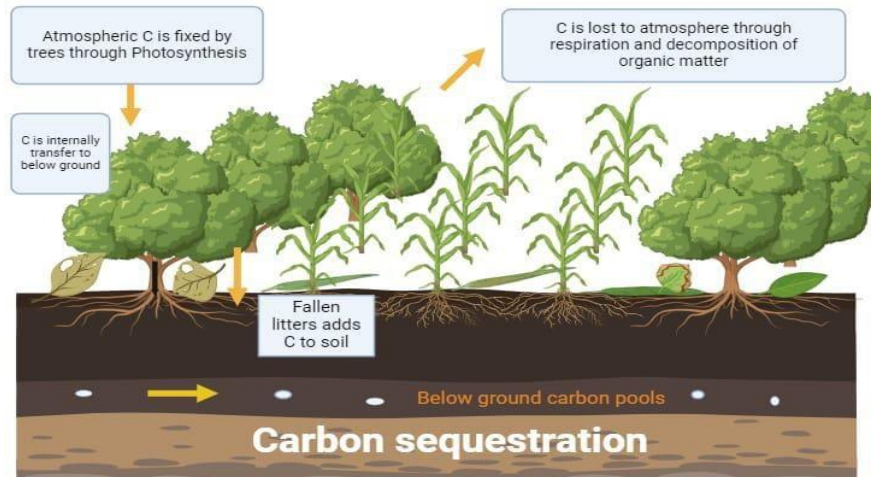

**Fig.3Carbon storage in soil underthe Agri-Silviculture system**

Since these trees are versatile and may be used for shade, fodder, fuel wood, fruit, veggies, and medicinal purposes, cultivating them sporadically on farmlands has been a long-standing and mostly unchanged tradition in India. Several agroforestry methods sequester carbon above and below the ground as carbon stored in standing biomass and as an improvement of soil carbon plus root biomass. Early research on the possibility of storing carbon in agroforestry systems and other alternative land use systems for India predicted a possible sequestration of 68 - 228 milligrams of carbon per hectare or 25 tons of carbon per hectare across 96 million hectares of land(Kendall & Pimentel, 1994). Up to 30 cm of soil, agroforestry can store around 83.6 tons of carbon per ha of carbon, which is 26% more than what can be grown in the plains of Haryana. Nonetheless, the size of the enterprise and the ultimate application of the wood would determine the amount of carbon sequestered by forestry operations. To manage runoff and soil erosion, reduced water, soil, organic matter, and nutrient losses are possible via agroforestry. It can prevent the growth of soil toxicities, salinisation, and acidification and help restore contaminated soils using trees(Montagnini & Nair, 2004).

The kind of agroforestry system significantly impacts the source or sink role of the trees, as evidenced by abundant data. Agrisilvipastoral systems, on the other hand, may be generators of greenhouse gases, whereas agrisilvicultural systems, in which crops and trees are produced together, are net sinks of $CO_2$. Though few field studies have supported these theories, agroforestry systems' capacity to sequester carbon has been demonstrated theoretically. The inconsistency of methods and the intrinsic heterogeneity in agroforestry system estimations of potential carbon storage have complicated comparisons(Jose, 2009)**.** Benefits of agroforestry systems include direct near-term carbon storage in trees and soils and the ability to offset immediate greenhouse gas emissions linked to deforestation and changing agriculture, according to assessments of national and global terrestrial $CO_2$sinks. It has been calculated that agroforestry on 96 million hectares of land in India has an average sequestration potential(Sathaye & Ravindranath, 1998)**.** Due to root penetration that prevents tillage, the tree species in the agroforestry system can store carbon in deeper soil layers for long-term storage. This is made possible by their ability to capture the most carbon through photosynthesis while lowering respiration rates and encouraging rapid development(Lorenz & Lal, 2014)**.** Farmlands are more carbon-sequestering than other land-use systems because they have more woody components(Sureshbhai et al., 2017)**.** Thus, it would seem that the agroforestry system is preferable for protecting biodiversity while offering social and economic benefits. Agroforestry systems vary widely in their capacity to store carbon, and the kind, structure, and function of the system, in addition to the surrounding area's environmental and socioeconomic aspects, all significantly impact the quantity of carbon stored. Fig.3may best illustratecombining silviculture with agriculture practices to improve soil carbon storage. This approach offers several advantages, including enhanced carbon sequestration, by carefully planting trees alongside crops.Two further factors that might affect the amount of carbon stored in an agroforestry system are tree species and system management.

Table 2 contains several more tree species and their carbon sequestration rate, as addressed by various writers. Carbon sequestration in soil under agroforestry systems has significant scientific and practical implications for climate change mitigation and sustainable agriculture. Agroforestry is the intentional incorporation of trees and shrubs into agricultural landscapes, and it has been acknowledged as an effective method of boosting soil carbon sequestration. Trees in agroforestry systems boost organic matter input through leaf litter and root biomass, which promotes soil carbon accumulation. Furthermore, the interaction between tree roots and soil microorganisms enhances organic carbon stabilisation in the soil matrix. This process mitigates atmospheric carbon dioxide levels and promotes soil health, water retention and nutrient cycling.

**Table 2: Carbon storage potential under different tree species**

| Tree species | Carbon sequestration rate (t/ha/year) | References |
|---|---|---|
| *Cocos nucifera* | ~2.61 | (Mina et al., 2023) |
| *Heveabrasiliensis* | ~16.61 | (Mina et al., 2023) |
| *Allocasuarin averticillata* | 3.09 | (Neumann et al., 2011) |
| *Corymbia maculate* | 3.88 | (Neumann et al., 2011) |
| *Eucalyptus camaldulensis* | 4.05 | (Neumann et al., 2011) |
| *Acacia nilotica* | 2.81 | (Kaur et al., 2002) |
| *Gmelinaar Korea* | 3.95 | (Negi et al., 1990) |
| *Albizia procera* | 3.70 | (Newaj & Dhyani, 2008) |
| *Albizia amara* | 1.00 | (Rai et al., 2000) |

Table 2 summarises the carbon storage potential of numerous tree species, providing a complete picture of their ability to sequester carbonwith quantitative data for each tree species.

## 4. Conclusion

This study emphasises that management approaches and land use changes impact SOC dynamics, such as soil aggregation, $CO_2$ efflux, and SOC quality. Land use changes pose problems for global agriculture since they impact $CO_2$ emissions into the atmosphere and global warming. SOC is the foundation of all terrestrial life and the preservation of natural resources. Reducing $CO_2$ emissions into the atmosphere and increasing SOC sequestration is essential to reducing global warming. Regardless of soil type, climate, land use or management techniques, temperature and moisture are the two main variables that control SOC dynamics. Significant relationships exist between SOC stocks, fractions, $CO_2$ outflow, temperature, and moisture. Compared to a baseline height, it has been demonstrated that a rise in altitude increases the SOC concentration by more than 5%. Site-specific factors such as temperature, soil characteristics, and plant species affect how quickly organic carbon builds up in soil. Soil carbon sequestration can occur when degraded or barren soils are converted to forests or permanent plants. A certain amount of SOC is released into the atmosphere when forest land is converted to croplands. Conversely, forest land converted to grasslands has a higher capacity to store carbon than farmland. Identifying appropriate land use and management techniques to reduce climate change by increasing soil carbon sequestration is crucial**.** When suggested management measures are adopted, the annual average rate of soil organic carbon sequestration generally surpasses the soil organic carbon depletion rate when land use changes to agriculture.

## 5. Way forward

Many of the management alternatives improve the overall sustainability of current agricultural systems increased the carbon content based on the economical value. Farming approaches that maximise soil organic carbon storage while preserving yield should be supported. Even though many mitigation options in the agricultural sector have numerous co-benefits in terms of food security, environmental sustainability, and farm profitability, economical feasibility in terms of grain yield, which encourages the adoption of best management practices, ongoing efforts should be made to evaluate different agriculture technologies with high sequestration potential and low global warming potential without compromising yield.

References

Ab, T., Gl, & Kulig, B. (2008). Effect of mulch and tillage system on soil porosity under wheat. *Soil tillage research*, 99, 169-178.

Adhikari, K., & Hartemink, A. E. (2016). Linking soils to ecosystem services—A global review. *Geoderma*, *262*, 101-111.

Albrecht, A., & Kandji, S. T. (2003). Carbon sequestration in tropical agroforestry systems. *Agriculture, ecosystem and environment.*, 99:15.27.

Amundson, R., & Biardeau, L. (2018). Soil carbon sequestration is an elusive climate mitigation tool. *Proceedings of the National Academy of Sciences of the United States of America.*

Anikwe, M., Mbah, C., Ezeaku, P., & Onyia. (2007). Tillage and plastic mulch effects on soil properties and growth and yeild of cocoyam on an ultisol in southeastern Nigeria. *Soil tillage research*, 93, 264-272.

Baker, J. M., Ochsner, T. E., Venterea, R. T., & Grifs, T. J. (2007). Tillage and soil carbon sequestration--What do we really know? *Agriculture, Ecosystem & Environment*, 118:111-115.

Bayer, C., Mielniczuk, J., Amado, T. J. C., Martin-Neto, L., & Fernandes, S. V. (2000). Organic matter storage in a sandy clay loam Acrisol affected by tillage and cropping systems in southern Brazil. *Soil and Tillage Research*, *54*(1-2), 101-109.

Bhatt, R., & Khera, K. (2006). Effect of tillage and mode of straw mulch application on soil erosion in the submontaneous tract of Punjab, India. *Soil tillage research*, 88, 107-115.

Bianchi, S. R., Miyazawa, M., Oliveira, E. L. d., & Pavan, M. A. (2008). Relationship between the mass of organic matter and carbon in soil. *Brazilian Archives of Biology and Technology*, *51*, 263-269.

Biederbeck, V. O., Campbell, C. A., and Zenter, R. P. (1984). Effect of crop rotation and fertilization on some biological properties of loam in southwester Saskatchewan. *Canadian journal of soil science.*, 64:355-367.

Carter, M. R. (2017). *Conservation tillage in temperate agroecosystems*. CRC Press.

Collins, H. P., Rasmunssen, P. E., and Dougkas, Jr., C. (1992). Crop rotation and residue management effects on soil organic and microbial dynamics. *Soil science society of America Journal.*, 56:783-788.

Deb, S. K., Shukla, M. K., Sharma, P., & Mexal, J. G. (2013). Soil water depletion in irrigated mature pecans under contrasting soil textures for arid southern New Mexico. *Irrigation science*, *31*, 69-85.

Dexter, A. R. (1991). Amelioration of soil by natural process. *Soil tillage research*, 20:87-100.

Dixon, R. K. (1995). Agroforestry systems: sources of sinks of greenhouse gases? *Agroforestry systems*, 31:99-116.

Don, A., Schumacher, J., & Freibauer, A. (2011). Impact of Tropical Land Use Change on soil organic carbon stocks-A Meta-Analysis. *Global Change Biology*, 17, 1658-1670.

Fonseca, L., Silva, V., Sá, J. C., Lima, V., Santos, G., & Silva, R. (2022). B Corp versus ISO 9001 and 14001 certifications: Aligned, or alternative paths, towards sustainable development? *Corporate social responsibility and Environmental Management*, *29*(3), 496-508.

Gajri, P., Arora, V., & Chaudhary, M. (1994). Maize growth responses to deep tillage, straw mulching and farmyard manure in coarse textured soils of NW India. *Soil use management*, 10, 15-19.

Giller, K. E., Witter, E., Corbeels, M., & Tittonell, P. (2009). Conservation agriculture and smallholder farming in Africa: The heretics view. *Field crops*, 114:123-124.

Griffin, E., Hoyle, F., & Murphy, D. (2013). Soil organic carbon. *Report card on sustainable natural resource use in agriculture'.(Department of Agriculture and Food, Western Australia: South Perth, W. Aust.).*

Gupta, R., Singh, A., Kumar, A., & Yadav, D. S. (2021). Impact of cow dung amendment on soil organic carbon, microbial biomass carbon and enzyme activities under long-term rice–wheat cropping system. *Gupta, R., Singh, A., Kumar, A., & Yadav, D. S. (2021). Impact of cow dung amendment on soil organic carbon, microbial biomass carbon and enzyme Journal of Environmental Management,*, 290, 112585.

Halvorson, A. D., Reule, C. A., & Follett, R. F. (1999). *Soil Science Society of America Journal*, 63:912-917.

Halvorson, A. D., Wienhold, B. J., & Black, A. L. (2002). *Soil Science Society of America Journal*, 66: 906-912.

Ingram, J. S. I., & Fernandez, E. C. M. (2001). Managing carbon sequestration in soils: concepts and terminology. *Agriculture ecosystem and environment*, 87:111-117.

Jackson, R. B. (2017). Warning signs for stabilizing CO2 emissions. *Environ Res Lett 12*.

Jarecki, M. K., & Lal, R. (2003). Crop management for soil carbon sequestration. *Critical Reviews in Plant Sciences*, *22*(6), 471-502.

Jastrow, J. (1996). Soil aggregate formation and the accrual of particulate and mineral associated organic matter. *Soil Biology and Biochemistry*, 28,665-676.

Jose, S. (2009). Agroforestry for ecosystem service and environmental benefits: An overview. *Agroforestry systems*, 76:71-10.

Kahlon, M. S., Lal, R., & Ann-Varughese, M. (2013). Twenty two years of tillage and mulching impacts on soil physical characteristics and carbon sequestration in Central Ohio. *Soil and Tillage Research*, *126*, 151-158.

Kassam, A., & Friedrich, T. (2012). *An ecologically sustainable approach to agricultural production intensification: Global perspective and developments*. http://journals.openedition.org/factsreports/1382

Kaur, B., Gupta, S., & Singh, G. (2002). Carbon storage and nitrogen cycling in silvopastoral systems on a sodic in northwestern India. *Agroforestry systems*, *54*, 21-29.

Kendall, H. W., & Pimentel, D. (1994). Constraints on the expansions of the global food supply. *Ambio*, 198-205.

Kimble, J. M., Follett, R. F., & Cole, C. V. (1998). *The potential of US cropland to sequester carbon and mitigate the greenhouse effect*. CRC Press.

Kucharik, C. J., Brye, K. R., Norman, J. M., Foley, J. A., Gower, S. T., & Bundy, L. G. (2001). Measurements and modeling of carbon and nitrogen cycling in agroecosystems of southern Wisconsin: Potential for SOC sequestration during the next 50 years. *Ecosystems*, 237-258.

Kumar, P., Gupta, P., Yadav, A., & Kanwar, P. (2020). Impact of farmyard amnure and nitrogen on soil organic carbon, microbial biomass and enzyme activities in maizr-wheat system. *Archives of Agronomy and soil science*, 66(62), 233-248.

Lal, R. (2001). Potential of desertification control to sequester carbon and mitigate the accelerated greenhouse effect. *Climate change*, 35-72.

Larney, F. J., & Kladivko, E. J. (1989). Soil strength properties under four tillage systems at three long term study sites in Indiana. *Soil science*, 1539-1545.

Lorenz, K., & Lal, R. (2014). Soil organic C sequestration in agroforestry systems.A review. *Agronomy for sustainable development.*, 34, 443-454.

Malaviya, M. K., Singh, K., & Jat, H. S. (2019). Malaviya, Impact of green manuring and crop residue incorporation on soil organic carbon and crop productivity in a rice–wheat cropping system. *Malaviya, M. K., Singh, K., & Jat, H. S. (2019). Impact of green manuring and crop residue incorSoil and Tillage Research,*, 185, 173-182.

Manivannan, S., Balamurugan, M., Parthasarathi, K., Gunasekaran, G., & Ranganathan, L. (2009). Effect of vermicompost on soil fertility and crop productivity-beans (Phaseolus vulgaris). *Journal of environmental biology*, *30*(2), 275-281.

Mina, U., Geetha, G., Sharma, R., & Singh., D. (2023). Comparative Assessment of Tree Carbon Sequestration Potential and soil carbon dynamica of major plantation crops and homestead agroforestry of kerala, In dia. *Anthropocene Science*, 5.

Montagnini, F., & Nair, P. (2004). Carbon sequestration:An underexploited environmental benefits of agroforestry systems. *Agroforestry systems*, 61:281-295.

Nath, C. P., Das, T. K., Rana, K. S., Bhattacharyya, R., Pathak, H., Paul, S., . . . Singh, S. B. (2017). Greenhouse gases emission, soil organic carbon and wheat yield as affected by tillage systems and nitrogen management practices. *Archives of Agronomy and Soil Science*, *63*(12), 1644-1660.

Negi, J., Bahuguna, V., & Sharma, D. (1990). Biomass production and distribution of nutrients in 20 years old teak (Tectona grandis) and gamar (Gmelina arborea) plantation [s] in Tripura. *Indian Forester*, *116*(9), 681-686.

Neumann, C. R., Hobbs, T. J., & Tucker., M. (2011). *Carbon sequestration and biomass production rates from agroforestry in lower rainfall zones of SA: Southern Murray-Darling Basin Region*.

Newaj, R., & Dhyani, S. (2008). Agroforestry systems for carbon sequestration: present status and scope. *Indian Journal of Agroforestry*, *10*(1).

Omay, A. B., Rice, C. W. Maddux, L. D., and Gordon,. (1997). Changes in soil microbial and chemical properties under long term crop rotation and fertilization. *Soil science society of America Journal.*, 61: 1672-1678.

Pacala, S., & Socolow, R., ,. (2004). Stabilizing wedges: solving the climate problem for the next 50 years with current technologies. *science*, 305, 968-972.

Pandey, D. Y. (2007). Multifunctional agroforestry systems in india. *Current science*, 92:455-463.

Poeplau, C., Don, A., Vesterdal, L., Leifeld, J., Wesemael, B., Van, Schumacher, J., & Gensior, A. (2011). Temporal Dynamics of soil organic carbon afterland use change in the temperate zone carbon responce function as a model approach. *Global Change Biology*, 17, 2415-2427.

Rai, P., Solanki, K. R., & Singh, U. P. (2000). Growth and biomass production of multipurpose tree species in natural grassland under semi arid condition. *Indian Journal of Agroforestry*, *2*(1&2).

Reeves, D. (1997). The role of soil organic matter in maintaining soil quality in continues cropping systems. *Soil till*, 43.

Sa, J., De Moraes, & Lal, R. (2009). Stratification ratio of soil organic matter pools as an indicator of carbon sequestration in tillage chronosequence on a Brazilian Oxisol. *Soil tillage research*, 103, 146-156.

Sainju, U. M., Singh, B. P., Whitehead, W. F., & Wang, S. (2006). Carbon supply and storage in tilled and nontilled soils as influenced by cover crops and nitrogen fertilization. *Journal of environmental quality*, *35*(4), 1507-1517.

Sanderman, J., Hengl, T., & Fiske, G. J. (2017). Soil carbon debt of 12000 years of human land use. *Proceedings of the national Academy of sciences*, 9575-9580.

Sarkar, S., & Singh, S. (2007). Interactive effect of tillage depth and mulch on soil temperature, productivity and water use pattern of rainfed barely. *Soil tillage research*, 92,79-86.

Sathaye, J. A., & Ravindranath, N. H. (1998). Climate change mitigation in the energy and forestry sectors of developing countries. *Annual review of environment and resources*.

Schlesinger, W. (2000). *Biogeochemistry*, 7-20.

Schlesinger, W. H., & Andrews, J. A. (2000). Soil respiration and the global carbon cycle. *Biogeochemistry*, *48*, 7-20.

Schroeder, P. (1993). Agorforestry systems: intergrated land use to store and conserve carbon. *Climate research*, 3:53-60.

Shirani, H., Hajabbasi, M. A., Afyuni, M., & Hemmat, A. (2002). Effects of farmyard manure and tillage systems on soil physical properties and corn yield in central Iran. *Soil and Tillage Research*, *68*(2), 101-108.

Shirani, H., Hajabbasi, M. A., Afyuni, M., and Hemmat,A. (2002). Effect of farmyard manure on tillage system on soil physical properties and corn yield in central Iran. *Soil tillage research*, 68: 101-108.

Shirley, D. W., & Teasdale, J. R. (1998). Influence of herbicide Application timing on corn production in a hairy vetch cover crop. *Journal of production agriculture*, 11, 121-125.

Singh, T. P., Varalakshmi, V., Ahluwalia, S.K.,. (2000). Carbon sequestration through farm forestry: case from India. *Indian Forester*, 126, 1257–1264.

Six, J., Elliott, E., & Paustian, k. (2000). Soil macroaggregate turn over and microaggregate formation: A mechanism for C sequestration under no-tillage agriculture. *Soil Biology and Biochemistry*, 32,2099-2013.

Skoien, S. (1993). Long term effects of crop rotation, manure and straw on soil aggregation. *Norwegian journal of Agricultural science*, 7:231-247.

Sohrabi, Y., Habibi, A., Mohammadi, K., Sohrabi, M., Hwidari, G., Khalesro, S., & Khalvandi, M. (2012). Effect of nitrogen fertilizerand foliar-applied iron fertilizer at various reproductive stageson yield component and chemical composition of soybean seed. *African Journal of Biotechnology*, 11, 9599-9605.

Sureshbhai, P. J., Thakur, N. S., Jha, S. K., & Kumar, V. (2017). Productivity and carbon sequestration under prevalent agroforestry systems in Navsari District, Gujarat,India. *International journal of current microbiology and applied science.*, 6, 3405–3422.

Uhlen, G., & Tveitnes, S. (1995). Effects of long-term crop rotation, fertilizers, farm manure and straw on soil productivity. *Nor J Agric Sci*, *9*, 143-161.

Wang, H., Wang, S., Yu, Q., Zhang, Y., Wang, R., Li, J., & Wang, X. (2020). No tillage increases soil organic carbon storage and decreases carbon dioxide emission in the crop residue returned farming system. *Environment*, 261, 110-261.

Zhang, Y., Yingqi Niu, and Tao Zhang. (2022). Introductory Chapter: The Overview of Recent Advances of Sustainable Waste Management.